\documentclass[arma]{svjour}
\usepackage[mathscr]{eucal}
\usepackage{amsmath,amssymb,amscd}
\usepackage{graphicx}
\usepackage{latexsym}
\def \beq{ \begin{equation} }
\def \eeq{\end{equation}}
\def\bq{\begin{eqnarray}}
\def\eq{\end{eqnarray}}
\def\bi{\begin{itemize}}
\def\ei{\end{itemize}}
\def\ba{\begin{array}}
\def\ea{\end{array}}
\def\R{{\bf R}}

\def \pd{\partial}

\def \q {\quad}

\def \R{{\mathbb R}}

\begin{document}
\title{Convex Four Body Central Configurations with Some Equal Masses}
\titlerunning{Central Configurations of the Four Body Problem}
\author{ Ernesto Perez-Chavela and Manuele Santoprete}
\date{Received: date / Revised version: date}




\maketitle
\begin{abstract}
We prove that  there is a unique  convex non-collinear central
configuration of the planar Newtonian four-body problem when two
equal masses are located at opposite vertices of a quadrilateral
and,  at most, only one of the remaining masses is larger than the
equal masses. Such central configuration posses a symmetry line and it is a kite shaped quadrilateral. 
 We also show that there is exactly one  convex
non-collinear central configuration when the opposite masses are
equal. Such central configuration also  posses a symmetry line and it is a rhombus.
\end{abstract}

\section{Introduction}
The Newtonian planar  $n$-body problem is the study of the dynamics
of $n$ point particles with masses $m_i\in{\mathbb R}^+$ and
positions $q_i\in{\mathbb R}^2$ ($i=1,\ldots,n$), moving according
to Newton's laws of motion: \beq m_i\ddot q_i=\frac{\partial
U}{\partial q_i}, \eeq where $U(q)$ is the Newtonian potential \beq
U(q)=\sum_{i<j}\frac{m_im_j}{r_{ij}} \eeq where
$r_{ij}=\|q_i-q_j\|$. Let  $q=(q_1,\ldots,q_n)\in {\mathbb R}^{2n}$
and let $M$ be the matrix
$\mathrm{diag}[m_1,m_1,m_2,m_2\ldots,m_n,m_n],$ then the equations
of motion can be written as follows:  \beq \ddot
q=M^{-1}\frac{\partial U}{\partial q}. \eeq

To study this problem, without any loss of generality, we can assume
the center of mass is fixed at the origin and  consider the  space
\[
\Omega_n=\{q=
(q_1,q_2,\ldots,q_n)\in\mathbb{R}^{2n}|\sum_{i=i}^nm_iq_i=0\}.
\]
Because the potential is singular when two particles have the same
position it is natural to assume that the configuration avoids the
set $\Delta=\bigcup_{i\leq j} \Delta_{ij}$ where
\[
\Delta_{ij}=\{(q_1,q_2,\ldots,q_n)\in\mathbb{R}^{2n}|q_i=q_j\}.
\]
The set $\Omega_n\setminus \Delta$ is called {\it the configuration
space}.

\begin{definition}
A configuration $q\in \Omega_n\setminus \Delta$ is called a {\it
central configuration} (c.c.) if there is some constant $\lambda$
such that \beq M^{-1}\frac{\partial U}{\partial q}=\lambda q.
\label{cc} \eeq
\end{definition}
Equations (\ref{cc}) are invariant under rotation, dilatation and
reflection on the plane. Two central configurations are considered
equivalent if they are related by these symmetry operations.

The question of the existence and classification of central
configuration is a fascinating problem that dates back to the 18th
century. In 1767, Euler discovered the collinear c.c.'s. In 1772
Lagrange proved that, for any three arbitrary masses, the
equilateral triangle is a central configuration.

For the collinear $n$-body problem an exact count of the central
configurations of $n$ bodies was found by Moulton \cite{Moulton}
(see also \cite{smale70} for a modern proof). There is a unique
collinear relative equilibrium for any ordering of the masses so
there are $n!/2$ collinear equivalence classes.

The number of planar central configurations of the $n$--body problem
for an arbitrary given set of positive masses, has been estabilished
only for $n=3$: there are always five relative equilibria. Two of
these are Lagrange's equilateral triangles and the other three are
collinear c.c. discovered by Euler. Already in the four body problem
there is sufficient complexity to prevent a complete classification
of the non-collinear relative equilibria. In fact, an exact count is
known only for the equal masses case \cite{Albouy95,Albouy96} and
for certain cases where some of the masses are assumed sufficiently
small \cite{Xia91,Tien93}.

Even the finiteness of the central configurations is a very
difficult question. This conjecture was proposed by Chazy
\cite{Chazy} and Wintner \cite{Wintner} and  was listed by Smale as
problem number $6$  on his list of problems for this century
\cite{Smale98}.  The finiteness problem was  settled by A. Albouy
\cite{Albouy95,Albouy96} for the case of four equal masses and by
Marshall Hampton and Rick Moeckel \cite{Hampton06} for the general
four body problem.

Aside from these fundamental results very little else is known in
terms of the classification of c.c.'s for $n\geq 4$. One interesting
open problem concerning the classification of c.c.'s, recently
emphasized by A. Albouy and Y. Fu \cite{Albouy06}, is the following:
{\it Prove that, in the planar four-body problem,  there is exactly
one convex central configuration such that two given masses are not
adjacent ( i.e. they are on the same diagonal)}.

It is the scope of this paper to solve this conjecture in two
particular cases. A first step in this direction was done by Y. Long
and S. Sun \cite{Long}. They proved that any convex non-collinear
convex central configuration with masses $\delta>\alpha>0$, such
that the diagonal corresponding to the mass $\alpha$ is not shorter
than the one  corresponding to the mass $\delta$, must posses a
symmetry and therefore must be a kite. However, in their paper, they
ask whether there are asymmetryc c.c.'s when the diagonal
corresponding to the mass $\alpha$ is shorter than the other one. We
show that this is not possible.

The main result of the paper is an extension of the above result,
where we consider that only two of the masses are equal  and  at
most, only one of the remaining masses is larger than the equal
masses. We have the following

\begin{theorem}
Let $q=(q_1,q_2,q_3,q_4)\in \Omega_4$ be a convex non-collinear
central configuration  with masses $(\delta,\delta,\alpha,\beta)\in
(\R^+)^4$. Suppose that the equal masses  are opposite vertices and
that $\alpha\leq \delta$ or $\beta \leq \delta$. Then the
configuration $q$ must posses a symmetry, it is unique and forms a
kite. \label{thmain1}
\end{theorem}

The uniqueness of the kite central configuration, in the hypothesis
of the above theorem  was proved by E. Leandro in \cite{Leandro}. We
therefore have the following

\begin{corollary}
Under the hypothesis of Theorem \ref{thmain1} there is exactly one
central configuration.
\end{corollary}

In particular, in the case $\alpha=\beta$ we prove the following 

\begin{theorem}
Let $q=(q_1,q_2,q_3,q_4)\in \Omega_4$ be a convex non-collinear
central configuration  with masses $(\delta,\delta,\alpha,\alpha)
\in (\R^+)^4$. Suppose that the equal masses  are opposite vertices
then the configuration $q$ must posses a symmetry and forms a
rhombus. \label{thmain2}
\end{theorem}

This theorem completely  answers the question of Y. Long and S. Sun \cite{Long}.
The uniqueness of the rhombus central configuration, in the
hypothesis of the above theorem, is easy to prove (see for example
\cite{Long} for a simple proof). We therefore have the following

\begin{corollary}
Under the hypothesis of Theorem \ref{thmain2} there is exactly one
central configuration.
\end{corollary}

In the next section we give some basic results and settings. In
Section 3 we prove Theorem \ref{thmain1}. In Section 4 we prove
Theorem \ref{thmain2}. In the Appendix we list different ways to
write the equations of the balanced configurations of A. Albouy and
A. Chenciner \cite{Albouy1998}.
\begin{figure}[t]
  \begin{center}
    \resizebox{!}{5.5cm}{\includegraphics{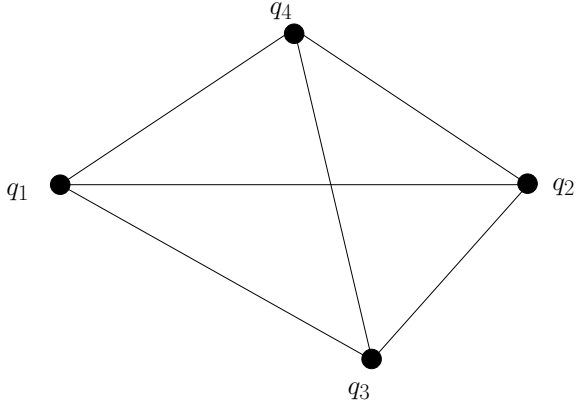}} \label{convex}
    \caption{A convex configuration of four masses.}
  \end{center}
\end{figure}

\section*{Acknowledgments}
The authors would like to thank A. Albouy and to an anonymous
referee for their comments and suggestions regarding this work.
Ernesto P\'erez-Chavela was supported by  CONACYT M\'exico, grant
no. 47768.

\section{Preliminaries}
Firstly observe that if $(q_1,q_2,q_3,q_4)\in \Omega_4$ is a central
configuration with parameter $\lambda$ and positive masses
$(m_1,m_2,m_3,m_4)$ then, for every $\eta>0$,
$(\frac{q_1}{\sqrt[3]{\eta}},\frac{q_2}{\sqrt[3]{\eta}},\frac{q_4}{\sqrt[3]{\eta}},
\frac{q_4}{\sqrt[3]{\eta}})\in \Omega_4$ is the same  central
configuration with masses
$(\frac{m_1}{\eta},\frac{m_2}{\eta},\frac{m_3}{\eta},\frac{m_4}{\eta})$
and the same value of $\lambda$. So, without loss of generality we
suppose $\delta=1$, and we consider the planar 4-body problem with
masses
\[
m_1=m_2=1,\q m_3=\alpha,\quad m_4=\beta.
\]

In this paper we use Dziobeck coordinates, that we  describe below
(see \cite{Albouy95} and \cite{Long} for more details).  Let \[
a=r_{12}^2,~ b=r_{13}^2,~ c=r_{14}^2,d=r_{23}^2,~ e=r_{24}^2,~
f=r_{34}^2.
\]

For $1\leq i\leq4$ let $|\Delta_i|$ be the area of the sub-triangle
formed by the remaining three vertices of the configuration $q$ when
deleting the point $q_i$. As in \cite{Long}, we define the oriented
areas of these sub-triangles of the convex non-collinear
configuration $q$ by
\[\Delta_1=-|\Delta_1|, \quad \Delta_2=-|\Delta_2|, \quad
\Delta_3=|\Delta_3|, \quad \Delta_4=|\Delta_4|\] when the masses are
opposite vertices of a quadrilateral. The $\Delta_i$ above  satisfy
the equation \beq \Delta_1+\Delta_2+\Delta_3+\Delta_4=0
\label{eqdeltas} \eeq

The Cayley determinant \beq S=\begin{vmatrix}
0&1&1&1&1\\
1&0&a&b&c\\
1&a&0&d&e\\
1&b&d&0&f\\
1&c&d&f&0
  \end{vmatrix}
\eeq satisfies $S=0$. In 1900 Dziobek proved that \beq \frac{\pd
S}{\pd r_{ij}^2}=32 \Delta_i\Delta_j \qquad \forall i\neq j. \eeq
Let $\psi(s)=s^{-1/2}$ for $s>0$. Then the potential function and
the moment of inertia are given by \beq U=\sum_{1\leq i<j\leq 4}
m_im_j\psi(r_{ij}^2)\label{pot} \eeq and \beq
I=\frac{1}{m'}\sum_{1\leq i<j\leq 4} m_im_j r_{ij}^2 \eeq
respectively, where $m'=\sum_{i=1}^4 m_i$. Using Lagrange
multipliers, Dziobek characterized the central configurations of
four bodies as the extrema of
\[
U-\lambda S-\mu(I-I_0)
\]
as a function of $\lambda,\mu,r_{12},\ldots,r_{34}$, where $\lambda$
and $\mu$ are Lagrange multipliers and $I_0$ is a fixed moment of
inertia. Thus, for any $i,j$ with $1\leq i<j\leq 4$, the central
configurations satisfy \beq \frac{\pd U}{\pd r_{ij}^2}=\lambda
\frac{\pd S}{\pd r_{ij}^2}+\mu \frac{\pd I}{\pd
r_{ij}^2}\label{eqdz}, \eeq and from (\ref{pot}), we also have
\[ \frac{\pd U}{\pd r_{ij}^2}=m_im_j \psi'(r_{ij}^2), \] where
$\psi'(s)$ denotes the derivative of the function $\psi(s)$ with
respect to $s$ and
\[\frac{\pd I}{\pd r_{ij}^2}=\frac{m_im_j}{m'}.\]
Consequently, equation (\ref{eqdz}) becomes \beq m_im_j
\psi'(r_{ij}^2)=32\lambda \Delta_i\Delta_j +\frac{m_im_j\mu}{m'}.
\label{eqdz1} \eeq Therefore, using our mass convention, the
equations for the central configurations are:

\begin{subequations}
\begin{eqnarray}
\psi'(r_{12}^2)=& \nu \Delta_1\Delta_2+\xi \label{subeqa}\\
\psi'(r_{13}^2)=& \frac\nu\alpha  \Delta_1\Delta_3+\xi\\
\psi'(r_{14}^2)=&\frac \nu \beta  \Delta_1\Delta_4+\xi\\
\psi'(r_{23}^2)=&\frac {\nu}{\alpha} \Delta_2\Delta_3+\xi\\
\psi'(r_{24}^2)=&\frac \nu \beta \Delta_2\Delta_4+\xi\\
\psi'(r_{34}^2)=&\frac{\nu}{\alpha\beta} \Delta_3\Delta_4+\xi
\label{subeqf}
\end{eqnarray}
\end{subequations}
where $\nu=32\lambda$ and $\xi=\frac{\mu}{m'}$. Moreover, there are
implicit relations between the $r_{ij}^2$ and  the $\Delta_i$: \beq
t_k=\sum_{i=1}^4 \Delta_i r_{ik}^2, \quad t_1=t_2=t_3=t_4.
\label{eqt}\eeq Using the implicit relations above we can prove the
following two Lemmas due to A. Albouy \cite{Albouy04}
\begin{lemma}
For a  central configuration, the corresponding $\nu$ in the
equations (\ref{subeqa})-(\ref{subeqf}) is positive. \label{mac}
\end{lemma}
\begin{lemma} The following inequality holds:
\beq
(\frac{\Delta_i}{m_i}-\frac{\Delta_j}{m_j})(\Delta_i-\Delta_j)\geq
0. \label{eqal} \eeq Consequently  $\Delta_i>\Delta_j$  if and only
if  $\frac{\Delta_i}{m_i}>\frac{\Delta_j}{m_j}$. \label{albouy}
\end{lemma}

We now prove Lemma \ref{mac} and  Lemma \ref{albouy}. From
(\ref{eqt}) we deduce
\[0=t_i-t_j=r_{ij}^2(\Delta_j-\Delta_i)+\sum_k\Delta_k(r_{ik}^2-r_{jk}^2)\]
and
\[ 0=\left(\frac{\Delta_i}{m_i}-\frac{\Delta_j}{m_j}\right)
(\Delta_j-\Delta_i)r_{ij}^2+m_k\sum_k\left(\frac{\Delta_i\Delta_k}{m_i
m_k}-\frac{\Delta_j\Delta_k}{m_jm_k}\right)(r_{ik}^2-r_{jk}^2).
\]
Using equation (\ref{eqdz1}) we get
\[
0=\left(\frac{\Delta_i}{m_i}-\frac{\Delta_j}{m_j}\right)
(\Delta_j-\Delta_i)r_{ij}^2+\frac{m_k}{32\lambda}
\sum_k(\psi'(r_{ik}^2)-\psi'(r_{jk}^2))(r_{ik}^2-r_{jk}^2).
\]
Since $\psi'(s)$ is a monotone increasing function of $s$
$(\psi'(r_{ik}^2)-\psi'(r_{jk}^2))(r_{ik}^2-r_{jk}^2)\geq0$. Thus
\[
\lambda(\frac{\Delta_i}{m_i}-\frac{\Delta_j}{m_j})(\Delta_i-\Delta_j)\geq
0
\]
Let us choose the index $i$ corresponding to the smallest
$\Delta_i$, and $j$ corresponding to the greatest $\Delta_j$. We
have $\Delta_i<0<\Delta_j$, because $\sum\Delta_k=0$.  Moreover  if
$\lambda=0$  all the edges are equal, but this is geometrically
impossible, thus $\lambda>0$, $\nu>0$. This concludes Lemma
\ref{mac}. Moreover
\[
(\frac{\Delta_i}{m_i}-\frac{\Delta_j}{m_j})(\Delta_i-\Delta_j)\geq
0. \quad\mbox{ for any }i,j
\]
this concludes the proof of Lemma \ref{albouy}.

Let \beq A=\psi'(a), B=\psi'(b),\ldots, F=\psi'(f). \eeq From
(\ref{eqt}) one can extract some weaker identities \beq
Q_{ijk}=\begin{vmatrix}
1&1&1\\
t_i&t_j&t_k\\
\frac{\Delta_i}{m_i}&\frac{\Delta_j}{m_j}&\frac{\Delta_k}{m_k}
        \end{vmatrix}=0.\label{eqq}
\eeq Of course $Q_{ijk}=0$ if $t_i=t_j=t_k$. But if all the
$Q_{ijk}$ are zero one can only deduce that
$t_i=\eta\Delta_i/m_i+\delta$ for some $(\eta,\delta)\in {\mathbb
R}^2$ and for all $i$. For the four body problem  equations
(\ref{eqq}), using the fact that
$\Delta_1+\Delta_2+\Delta_3+\Delta_4=0$, after some tedious
computations we obtain
\beq
\begin{vmatrix}
1&1&1\\
f-e-d&\alpha(e-d-f)&\beta(d-f-e)\\
F&E&D
\end{vmatrix}=\begin{vmatrix}
1&1&1\\
a+f&b+e&c+d\\
A&B&C
\end{vmatrix}\label{eqconf1}
\eeq
\beq
\begin{vmatrix}
1&1&1\\
f-c-b&\beta(b-f-c)&\alpha(c-b-f)\\
F&B&C
\end{vmatrix}=\begin{vmatrix}
1&1&1\\
a+f&b+e&c+d\\
A&E&D
\end{vmatrix}\label{eqconf2}
\eeq
\beq
\begin{vmatrix}
1&1&1\\
\beta(a-e-e)&e-c-a&c-a-e\\
A&E&C
\end{vmatrix}=\alpha\begin{vmatrix}
1&1&1\\
a+f&b+e&c+d\\
F&B&D
\end{vmatrix}\label{eqconf3}
\eeq
\beq
\begin{vmatrix}
1&1&1\\
\alpha(a-d-b)&b-a-d&d-b-a\\
A&B&D
\end{vmatrix}=\beta\begin{vmatrix}
1&1&1\\
a+f&b+e&c+d\\
F&E&C
\end{vmatrix}\label{eqconf4}
\eeq

These  are the equations of the {\it balanced configurations}
(configuration \'equilibr\'ee) due to A. Albouy and A. Chenciner
\cite{Albouy1998}. In the Appendix we present other ways to write
the above identities.

Observe that  the determinant \beq d(u,v,w;U,V,W)= \begin{vmatrix}
1&1&1\\
u&v&w\\
U&V&W
 \label{triangle}               \end{vmatrix}\eeq
has a beautiful geometrical interpretation. In fact $d(u,v,w;U,V,W)$
is the oriented area of the triangle of vertices $(u,U), (v,V),
(w,W)$ (see \cite{Uspensky}) where the sign is determined by the
following Lemma:
\begin{lemma}
Let \[ V'=\frac{v-w}{u-w}U+\frac{u-v}{u-w}W.
\]
\begin{enumerate}
\item the following holds:
\[ d(u,v,w;U,V,W)=(u-w)(V-V'). \]
\item The determinant $d(u,v,w;U,V,W)>0$ provided $u>v>w$ and $V>V'$,
i.e. the point $(v,V)$ is located strictly above the line passing
through $(w,W)$ and $(u,U)$.
\item The determinant $d(u,v,w;U,V,W)<0$  provided $u>v>w$ and $V<V'$,
i.e. the point $(v,V)$ is located strictly below the line passing
through $(w,W)$ and $(u,U)$.
\item Let $g:(0,+\infty)\rightarrow(0,\infty)$ be a strictly concave function.
Suppose $u>v>w$. Then $d(u,v,w,g(u),g(v),g(w))>0$.
\end{enumerate}\label{lemmasign}
\end{lemma}
Lemma \ref{lemmasign} will  be useful in proving the main results of
this paper. A proof can be found in \cite{Long}.

\section{Proof of Theorem \ref{thmain1}}
Observe that, in order to have symmetry, under the hypotesis of
Theorem \ref{thmain1} the following inequality must hold
   \beq\Delta_1=\Delta_2 .\label{eqsymm}\eeq

Note that $\Delta_3/\alpha=\Delta_4/\beta$ or $\Delta_3=\Delta_4$
only if one is in a symmetric configurations with $\alpha=\beta$
(see \cite{Albouy,Albouy04} for more details). In this section we
assume $\beta\neq \alpha$. To show that a configuration is
symmetric, i.e. that (\ref{eqsymm}) holds, one can assume that
\[
\Delta_1\neq\Delta_2\quad \mbox{and}\quad \Delta_3\neq\Delta_4.
\]
and then derive a contradiction. This is the strategy of the proof.
If we assume $\Delta_1\neq\Delta_2\quad \mbox{and}\quad
\Delta_3\neq\Delta_4 $  we have four cases
 \beq
 \begin{split}
 &\mbox{\bf (a) } \Delta_1<\Delta_2<0<\Delta_3<\Delta_4\\
 &\mbox{\bf (b) } \Delta_2<\Delta_1<0<\Delta_3<\Delta_4\\
 &\mbox{\bf (c) } \Delta_1<\Delta_2<0<\Delta_4<\Delta_3\\
 &\mbox{\bf (d) } \Delta_2<\Delta_1<0<\Delta_4<\Delta_3.
\end{split}\label{cases}
 \eeq

Moreover the mutual distances satisfy some geometrical inequalities.
We have then
 \begin{lemma}
 The following inequalities hold
 \begin{subequations}\begin{eqnarray}
    &c< \min\{b,e\}\leq \max\{b,e\}<d<\min\{a,f\}\leq \max\{a,f\}\label{eqsub1a}\\
   &e< \min\{c,d\}\leq \max\{c,d\}<b<\min\{a,f\}\leq \max\{a,f\} \label{eqsub1b}\\
   &b< \min\{c,d\}\leq \max\{c,d\}<e<\min\{a,f\}\leq \max\{a,f\}\label{eqsub1c}\\
   &d< \min\{b,e\}\leq \max\{b,e\}<c<\min\{a,f\}\leq \max\{a,f\}\label{eqsub1d}
    \end{eqnarray}
 \end{subequations}
 in the cases a,b,c and d respectively.
 \label{lemmaineq}\end{lemma}
\begin{proof}
 
 We prove (\ref{eqsub1a}), the other cases are similar. Case (a) has
3 possible subcases.

\begin{itemize}
\item {Subcase 1.} $\Delta_1+\Delta_4=0$. In this case, by
equation (\ref{eqdeltas}) we obtain
\[\Delta_1=-\Delta_4,\quad \Delta_2=-\Delta_3.
\]

Thus we have
\[\Delta_1\Delta_4<\Delta_1\Delta_3={\Delta_2\Delta_4}<
\Delta_2\Delta_3<0<\Delta_1\Delta_2=\Delta_3\Delta_4.\]

Moreover by Lemma \ref{mac} and Lemma \ref{albouy} we have

\[\begin{split}\frac{\nu}{\beta}\Delta_1\Delta_4<
\min\{\frac{\nu}{\alpha}\Delta_1\Delta_3,\frac{\nu}{\beta}
\Delta_2\Delta_4\}\leq\max\{\frac{\nu}{\alpha}\Delta_1\Delta_3,
\frac{\nu}{\beta}\Delta_2\Delta_4\}\\ <
\frac{\nu}{\alpha}\Delta_2\Delta_3<0<\nu\Delta_1\Delta_2<
\frac{\nu}{\alpha\beta}\Delta_3\Delta_4.\end{split}\]

Therefore by equations (\ref{subeqa})-(\ref{subeqf}) and the
monotonicity of $\psi'(s)$ we have
 \beq
 c< \min\{b,e\}\leq \max\{b,e\}<d<a<f.
 \eeq

 \item Subcase 2. $\Delta_1+\Delta_4<0$.

 In this case, by equation (\ref{eqdeltas}) we have $\Delta_2+\Delta_3>0$.
 Using equation (\ref{eqdeltas}) yields
\[\begin{split}
\Delta_2\Delta_4-\Delta_1\Delta_3=\Delta_2\Delta_4+\Delta_1(\Delta_1+\Delta_2+\Delta_4)\\
=(\Delta_2+\Delta_1)(\Delta_4+\Delta_1)>0.
\end{split}\]
Similarly
 \beq\begin{split}
\Delta_3\Delta_4-\Delta_1\Delta_2&=\Delta_3\Delta_4+\Delta_1(\Delta_2+\Delta_3+\Delta_4)\\
&=(\Delta_1+\Delta_3)(\Delta_1+\Delta_4)>0\\
\end{split}\eeq
 since $\Delta_1+\Delta_4<0$ together with $\Delta_1<\Delta_2<0<\Delta_3<\Delta_4$ implies $\Delta_1+\Delta_3<0$.
 Consequently we have
 \[\Delta_1\Delta_4<\Delta_1\Delta_3<{\Delta_2\Delta_4}<\Delta_2\Delta_3<0<\Delta_1\Delta_2<\Delta_3\Delta_4\]
 and using  Lemma \ref{mac} and Lemma \ref{albouy} we obtain
  \[\begin{split}\frac{\nu}{\beta}\Delta_1\Delta_4<\min\{\frac{\nu}{\alpha}\Delta_1\Delta_3,\frac{\nu}{\beta}\Delta_2\Delta_4\}\leq\max\{\frac{\nu}{\alpha}\Delta_1\Delta_3,\frac{\nu}{\beta}\Delta_2\Delta_4\}\\ <\frac{\nu}{\alpha}\Delta_2\Delta_3<0<\nu\Delta_1\Delta_2<\frac{\nu}{\alpha\beta}\Delta_3\Delta_4.\end{split}\]
  Thus, as in Subcase 1, we get
 \beq
 c< \min\{b,e\}\leq \max\{b,e\}<d<a<f.
 \eeq

\item Subcase 3: $\Delta_1+\Delta_4>0$.

In this case   equation (\ref{eqdeltas}) implies that
$\Delta_2+\Delta_3<0$. Hence

 \[\begin{split}
\Delta_2\Delta_4-\Delta_1\Delta_3
=(\Delta_2+\Delta_1)(\Delta_4+\Delta_1)<0
\end{split}\]
and
\[\Delta_1\Delta_4<\Delta_2\Delta_4<{\Delta_1\Delta_3}<\Delta_2\Delta_3.\]
Using  Lemma \ref{mac} and Lemma \ref{albouy} we find
  \[\begin{split}\frac{\nu}{\beta}\Delta_1\Delta_4<
  \min\{\frac{\nu}{\alpha}\Delta_1\Delta_3,\frac{\nu}{\beta}\Delta_2\Delta_4\}
  \leq\max\{\frac{\nu}{\alpha}\Delta_1\Delta_3,\frac{\nu}{\beta}\Delta_2\Delta_4\}\\
   <\frac{\nu}{\alpha}\Delta_2\Delta_3<0<\nu\Delta_1\Delta_2 \end{split}\]
  and thus
  \[c<\min\{b,e\}\leq\max\{b,e\}<d<\min\{a,f\}\leq \max\{a,f\}.\]
Therefore, in all three subcases inequality (\ref{eqsub1a}) holds.
This concludes the proof of the Lemma. \hfill $\Box$
 \end{itemize}
 \end{proof}

We can now continue  the proof of Theorem \ref{thmain1}.
From (\ref{eqconf4}), using the fundamental properties of the determinants, we deduce

\beq (\alpha+1)\begin{vmatrix}
1&1&1\\
a&b&d\\
A&B&D
 \end{vmatrix}-(\alpha-1)[A(d-b)+bB-dD]=\beta\begin{vmatrix}
1&1&1\\
f&e&c\\
F&E&C
 \end{vmatrix}+
\beta \begin{vmatrix}
1&1&1\\
a&b&d\\
F&E&C
 \end{vmatrix}.
 \label{eqbalanced4a}
\eeq

Now consider the cases (a) and (c). Then,  by equation
(\ref{eqsub1a}) and (\ref{eqsub1c}), Lemma \ref{lemmasign} (2) and
the concavity of the function $\psi'(s)$, we obtain \beq
\begin{vmatrix}
1&1&1\\
a&b&d\\
A&B&D
 \end{vmatrix}<0,\quad
\begin{vmatrix}
1&1&1\\
f&e&c\\
F&E&C
 \end{vmatrix}>0,\quad
 \begin{vmatrix}
1&1&1\\
a&b&d\\
F&E&C
 \end{vmatrix}>0.
\eeq

We also have that $d-b>0$, $A<0$ and $bB-dD<0$ since

\[bB-dD=\frac1 2\left ( \frac{\sqrt{b}-\sqrt{d}}{\sqrt{bd}}  \right )<0.\]

Consequently $-(\alpha-1)[A(d-b)+bB-dD]<0$ when $0<\alpha\leq 1$ and
the left hand side of (\ref{eqbalanced4a}) is negative. This
produces a contradiction since the right hand side of
(\ref{eqbalanced4a}) is positive for any value of $\beta>0$.

In the cases (b) and (d) we have

\beq
\begin{vmatrix}
1&1&1\\
a&b&d\\
A&B&D
 \end{vmatrix}>0,\quad
\begin{vmatrix}
1&1&1\\
f&e&c\\
F&E&C
 \end{vmatrix}<0,\quad
 \begin{vmatrix}
1&1&1\\
a&b&d\\
F&E&C
 \end{vmatrix}<0.
\eeq

We also have that $d-b<0$, $A<0$ and $bB-dD>0$.

Consequently $-(\alpha-1)[A(d-b)+bB-dD]>0$ when $0<\alpha\leq 1$ and
the left hand side of (\ref{eqbalanced4a}) is positive for any value
of $\beta>0$. This produces a contradiction since the right hand
side of (\ref{eqbalanced4a}) is negative.

 From (\ref{eqconf3}), using the fundamental properties of the determinants we deduce

\beq (\beta+1)\begin{vmatrix}
1&1&1\\
a&e&c\\
A&E&C
 \end{vmatrix}-(\beta-1)[A(c-e)+eE-cC]=\alpha\begin{vmatrix}
1&1&1\\
a&e&c\\
F&B&D
 \end{vmatrix}+
\alpha \begin{vmatrix}
1&1&1\\
f&b&d\\
F&B&D
 \end{vmatrix}.
 \label{eqbalanced3a}
\eeq

Now consider the cases (a) and (c). Then we find

\beq
\begin{vmatrix}
1&1&1\\
a&e&c\\
A&E&C
 \end{vmatrix}>0,\quad
\begin{vmatrix}
1&1&1\\
a&e&c\\
F&B&D
 \end{vmatrix}<0,\quad
 \begin{vmatrix}
1&1&1\\
f&b&d\\
F&B&D
 \end{vmatrix}<0.
\eeq

We also have that $c-e<0$, $A<0$ and $eE-cC>0$ since

\[eE-cC=\frac1 2\left ( \frac{\sqrt{e}-\sqrt{c}}{\sqrt{ec}}  \right )>0.\]

Consequently $-(\beta-1)[A(c-e)+eE-cC]>0$ when $0<\beta\leq 1$ and
the left hand side of (\ref{eqbalanced4a}) is negative for any value
of $\alpha>0$. This produces a contradiction since the right hand
side of (\ref{eqbalanced4a}) is positive.

In the cases (b) and (d) we have \beq
\begin{vmatrix}
1&1&1\\
a&e&c\\
A&E&C
 \end{vmatrix}<0,\quad
\begin{vmatrix}
1&1&1\\
a&e&c\\
F&B&D
 \end{vmatrix}>0,\quad
 \begin{vmatrix}
1&1&1\\
f&b&d\\
F&B&D
 \end{vmatrix}>0.
\eeq

We also have that $c-e>0$, $A<0$ and $eE-cC<0$.

Consequently $-(\beta-1)[A(c-e)+eE-cC]>0$ when $0<\beta\leq 1$ and
the left hand side of (\ref{eqbalanced3a}) is negative for any value
of $\alpha>0$. This produces a contradiction since the right hand
side of (\ref{eqbalanced3a}) is positive.

This proves that $\Delta_1=\Delta_2$  and thus it shows the
existence of a line of symmetry.

We now show that the configuration must be a kite
\begin{lemma}
If $\Delta_1=\Delta_2$ the quadrilateral $q$ is a kite.
\label{lemmakitea}
\end{lemma}

\begin{proof}
 
 By equations (\ref{subeqa}-\ref{subeqf}) of the central
configurations we have \beq \psi'(r_{13}^2)=\frac \nu\alpha
\Delta_1\Delta_3+\xi=\frac \nu\alpha
\Delta_2\Delta_3+\xi=\psi'(r_{23}^2). \eeq Since $\psi'(s)$ is a
monotone increasing function of $s$ we obtain \beq r_{13}=r_{23}.
\eeq Similarly \beq \psi'(r_{14}^2)=\frac \nu\beta
\Delta_1\Delta_4+\xi=\frac \nu\beta
\Delta_2\Delta_4+\xi=\psi'(r_{24}^2). \eeq and thus \beq
r_{14}=r_{24}. \eeq 
Therefore  the quadrilateral is a kite. \hfill$\Box$
\end{proof}

This  concludes the proof of Theorem \ref{thmain1}.

\section{Proof of Theorem \ref{thmain2}}

In this case $\beta=\alpha$. Observe that, in order to prove Theorem \ref{thmain2} it is enough  to study the case $0<\alpha\leq \delta=1$. In fact the  case $\alpha\geq \delta=1$ can be obtained from the previous one just renaming the masses).

In order to prove the existence of a line of symmetry, in the configuration under consideration in this paper, we need to have either $\Delta_1=\Delta_2$ or $\Delta_3=\Delta_4$. 
As in Theorem \ref{thmain1} we can assume that $\Delta_1\neq \Delta_2$ and $\Delta_3\neq\Delta_4$  and prove the existence of the line of symmetry by contradiction. 
Even in this case one obtain the four cases in equation (\ref{cases}) and Lemma \ref{lemmaineq} holds. 

The reminder of the proof  follows directly from the one of Theorem \ref{thmain1} with $\beta=\alpha$, since as observed above one needs only to consider the case $\alpha\geq 1$.
  
The main reason to include this result as a Theorem is on one hand, that it completely answers the question formulated by Y. Long and S.
Sun in \cite{Long}, and on the other hand, that is not obvious from
the equations of the balanced configurations, or from Dziobek
equations that the kite central configuration obtained using Theorem \ref{thmain1} is a rhombus, but it follows from the uniqueness of
the rhombus central configuration and a result by E. Leandro
\cite{Leandro} that can be summarized as follows

\begin{lemma}
For any $\alpha>0$ and $\beta>0$, there exists a unique  central
configuration $q=(q_1,q_2,q_3,q_4)$ with masses $(1,1,\alpha,\beta)$
where $q_1$ and $q_2$ as well as $q_3$ and $q_4$ are located at the
opposite vertices of a kite shaped quadrilateral.
\end{lemma}

This  concludes the proof of Theorem \ref{thmain2}.

\section*{Appendix}
From (\ref{eqconf1}-\ref{eqconf4}), using the fundamental properties
of the determinants we deduce the following identities
\beq\begin{split}
-(1-\alpha)(f-e-d)(D-E)&+(\beta-\alpha)(d-f-e)(F-E)+
2\alpha\begin{vmatrix}
1&1&1\\
f&e&d\\
F&E&D
 \end{vmatrix}\\&=\begin{vmatrix}
1&1&1\\
a&b&c\\
A&B&C
 \end{vmatrix}+
 \begin{vmatrix}
1&1&1\\
f&e&d\\
A&B&C
 \end{vmatrix}.
\end{split}
\eeq

\beq\begin{split}
-(1-\alpha)(f-c-b)(C-B)&+(\beta-\alpha)(b-f-c)(C-F)
+2\alpha\begin{vmatrix}
1&1&1\\
f&b&c\\
F&B&C
 \end{vmatrix}\\&=\begin{vmatrix}
1&1&1\\
a&e&d\\
A&E&D
 \end{vmatrix}+
 \begin{vmatrix}
1&1&1\\
f&b&c\\
A&E&D
 \end{vmatrix}.
 \end{split}
\eeq

\beq -(\beta-1)(a-e-c)(C-E) +2\begin{vmatrix}
1&1&1\\
a&e&c\\
A&E&C
 \end{vmatrix}=\alpha\begin{vmatrix}
1&1&1\\
a&e&c\\
F&B&D
 \end{vmatrix}+
 \alpha\begin{vmatrix}
1&1&1\\
f&b&d\\
F&B&D
 \end{vmatrix},
\eeq
\beq -(\alpha-1)(a-d-b)(D-B) +2\begin{vmatrix}
1&1&1\\
a&b&d\\
A&B&D
 \end{vmatrix}=\beta\begin{vmatrix}
1&1&1\\
f&e&c\\
F&E&C
 \end{vmatrix}+
 \beta\begin{vmatrix}
1&1&1\\
a&b&d\\
F&E&C
 \end{vmatrix}.
\eeq
Note that if $\beta=\alpha$ the identities above reduce to the
expression of the balanced configurations used in \cite{Long}.

The equation of the balanced configurations can also be written in a
different way, that seems to be useful in certain problems (e.g. to
prove the main result of this paper, Theorem \ref{thmain1}). From
(\ref{eqconf1}-\ref{eqconf4}) we find
\beq\begin{split} (1+\beta)&\begin{vmatrix}
1&1&1\\
f&e&d\\
F&E&D
 \end{vmatrix}-(1-\alpha)[F(d-e)+eE-dD]-(\beta-\alpha)[D(e-f)+fF-eE]
 \\&=\begin{vmatrix}
1&1&1\\
a&b&c\\
A&B&C
 \end{vmatrix}+
 \begin{vmatrix}
1&1&1\\
f&e&d\\
A&B&C
 \end{vmatrix}.
\end{split}
\eeq

\beq\begin{split} (\beta+1)&\begin{vmatrix}
1&1&1\\
f&b&c\\
F&B&C
 \end{vmatrix}-(1-\alpha)[F(c-b)+bB-cC]-(\beta-\alpha)[B(f-c)+cC-fF]\\&=\begin{vmatrix}
1&1&1\\
a&e&d\\
A&E&D
 \end{vmatrix}+
 \begin{vmatrix}
1&1&1\\
f&b&c\\
A&E&D
 \end{vmatrix}.
 \end{split}
\eeq

\beq (\beta+1)\begin{vmatrix}
1&1&1\\
a&e&c\\
A&E&C
 \end{vmatrix}-(\beta-1)[A(c-e)+eE-cC]=\alpha\begin{vmatrix}
1&1&1\\
a&e&c\\
F&B&D
 \end{vmatrix}+
\alpha \begin{vmatrix}
1&1&1\\
f&b&d\\
F&B&D
 \end{vmatrix},
\eeq
\beq (\alpha+1)\begin{vmatrix}
1&1&1\\
a&b&d\\
A&B&D
 \end{vmatrix}-(\alpha-1)[A(d-b)+bB-dD]=\beta\begin{vmatrix}
1&1&1\\
f&e&c\\
F&E&C
 \end{vmatrix}+
\beta \begin{vmatrix}
1&1&1\\
a&b&d\\
F&E&C
 \end{vmatrix}.
\eeq


%
\address{
Departamento de Matem\'aticas\\
 UAM--Iztapalapa, A.P. 55--534\\
09340 Iztapalapa, Mexico, D.F., Mexico\\
email: {epc@xanum.uam.mx} \and
Department of Mathematics\\
Wilfrid Laurier University\\ 75 University Avenue  West, Waterloo, Canada\\
email: {msantoprete@wlu.ca}}

\end{document}